# MODELING OF ASTEROID SHAPES

## Kokorev A., Golubov O.

*V. N. Karazin Kharkiv National University*

**Summary:** In this article we consider different methods of modeling asteroid shapes, especially lightcurve inversion technique, and scattering laws used for it. We also introduce our program, which constructs lightcurves for a given asteroid shape model. It can be used to comparing shape model with observational data.

**Анотація:** У даній статті ми розглядаємо різні методи моделювання форми астероїдів, а особливо докладно — метод інверсії кривих освітлення, а також закони розсіяння світла, які для цього використовуються. Також пропонуємо нашу програму для побудови кривих освітлення за відомою модельною формою астероїда. Ця програма може бути використана для порівняння моделі з данними спостерігачів.

**Ключові слова:** астероїди, фотометричні методи, розсіяння світла.

**Аннотация:** В данной статье мы рассматриваем разные методы моделирования формы астероидов, особенно подробно — метод инверсии кривых блеска, а также законы рассеяния света, которые для этого используются. Также представляем нашу программу для построения кривых блеска для заданной модельной формы астероида. Эта программа может быть использована для сравнения модели с наблюдательными данными.

**Ключевые слова:** астероиды, фотометрические методы, рассеяние света.

## Introdaction

To determine diameter and shape of asteroids a lot of methods are used, such as: lightcurve inverse technique, star occultations by asteroids, radiolocation method, and finally direct on-site inspection. Let us look through some of them and consider deeply the first one.

1. Occultation method.

In this method a net of observers fixes motion of the asteroid shadow on the Earth's surface. Each observer measures the lapse of time during which star's brightness decreases due to asteroid passing. This time is in direct relation with the diameter and the shape of the asteroid [1]. For such kinds of observations CCD cameras are usually used, but direct visual observation is also possible. In the last case individual reaction time for each observer can cause significant systematic error.

First star occultation event was observed by P. Bjorklund and S. Muller in Sweden in 19 February 1958. That night asteroid Juno occultated star SAO 112328.

2. RADAR method.

Transmitter broadcasts a wave packet toward the asteroid. It is reflected from the surface and, according to the shape of the asteroid, comes back with some time delay. Furthermore, as far as asteroid rotates, we can detect the Doppler-effect from its different parts, and thus determine rotation period. Waves reflected from the part of the asteroid which rotates toward observer increase their frequency, and the ones from the part which rotates from observer decrease frequency [3].

To detect such phenomena radio telescopes with extremely high resolution are required. The errors should not exceed nanoseconds for time delay measurement, and hundredth of hertz for Doppler shift.

The great advantage of this method is its independence of passive light sources. The radar creates its own illumination and makes it possible to observe an asteroid regardless of the position of the sun.

### Lightcurve inversion technique

There are two ways of shape modeling [4]: octantoids based on spherical harmonics, and subdivision surfaces. The octanoid's starlike surface is given by the following parametrization:

where $a$, $b$ and $c$ are linear combinations of the spherical harmonic functions $Y^m_l(\theta, \varphi)$, with coefficients $a_{lm}$, $b_{lm}$ and $c_{lm}$, respectively. Then these coefficients are altered in

order to minimize $\chi^2$-function with certain regularization function. Let it be, for example:

$$\eta = \sum_{l,m} l(a_{lm}^2 + b_{lm}^2)$$

which reflects the shape's deviation from starlike form. After that we can expect to get a surface that will satisfy the observed lightcurves.

Second possibility is using subdivision surface. At first we should chose an initial set of vertices and corresponding triangles (control mesh), which have to fit observed lightcurve. Then apply to it the Loop scheme with a few steps. In this case another regularization function should be used:

Beside it, lightcurves can be applied to the determination of pole and rotation period of asteroid.

## Scattering law

A scattering law for lightcurve inversion model should be simple enough, i.e. we avoid too detailed physical parameters and try to focus on general photometric properties of the surface. For this purpose the combination of Lommel–Seeliger and Lambert law is usually chosen [2]. It is realistic enough, but more simple than e.g. Hapke model, which in addition gives unrealistic parameter values in inverse problems.

According to this law, the surface reflectance S as a function of the angle of incidence **i** and the angle of emergence **e** is then

$$S(\mu, \mu_0, \alpha) = f(\alpha) [S_{LS}(\mu, \mu_0) + cS_L(\mu, \mu_0)],$$

where $\mu_0 = \cos i$, $\mu = \cos e$, the Lommel-Seeliger single scattering term is $S_{LS}(\mu, \mu_0) = (\mu\mu_0)/(\mu + \mu_0)$, the Lambert multiple scattering term is $S_L(\mu, \mu_0) = \mu\mu_0$. $f(\alpha)$ determines dependence of reflectance on the phase angle $\alpha$.

## Building lightcurves

We are going to introduce you our program for checking compliance of asteroid shape model with real observation data. Four input parameters are required: the asteroid form itself, it's pole, rotation period (we took them from [6]), and ephemeris (could be taken from website [5]). The output is lightcurve for asteroid's luminosity and rendering of asteroid for specified location of observer and light source. In our program we use Lambert's scattering law as the simplest one.

1, Few words about general program's algorithm.

1. Transform input vectors to Sun and Earth $\mathbf{r}_{ecl}$ from geocentric ecliptic frame with origin translated to the asteroid to those ones in asteroid's own frame $\mathbf{r}_{ast}$ (z axis is co-directional with asteroid's axis of rotation):

$$\mathbf{r}_{ast} = M_z(\alpha) M_x(\delta) \, M_z(\varphi_0) \, \mathbf{r}_{ecl}$$

Here $M_i(\psi)$ are rotation matrix responsible for the rotation through angle $\psi$ in the positive direction about axis $i$. $\alpha$ and $\delta$ are the right ascension and declination of asteroid's pole in geocentric ecliptic frame, $\varphi_0$ is the initial phase of rotation.

2. Calculate normal unit vectors $\mathbf{n}_i$ for each facet of the asteroid.

3. Calculate asteroid's illumination detectable for the observer due to Lambert's scattering law:

$$E = A * \sum_i (\mathbf{n}_i \mathbf{e})(\mathbf{n}_i \mathbf{s}) * H(\mathbf{n}_i, \mathbf{e}) H(\mathbf{n}_i, \mathbf{s})$$

Here $\mathbf{n}_i$ and $S_i$ are unit normal vector and area of facet $i$ respectively. $\mathbf{e}$ and $\mathbf{s}$ are directions to the Earth and to the Sun in asteroid's own coordinate frame, $H$ is Heaviside step function introduced in order to eliminate unlighted and invisible for observer parts of the asteroid, $A$ is a normalization constant. Summation is over all facets.

4. Divide rotation period into a certain number of points and calculate illumination for all this positions to build the lightcurve.

5. Render asteroid directly from .obj file.

There are a few examples of output data for asteroid:

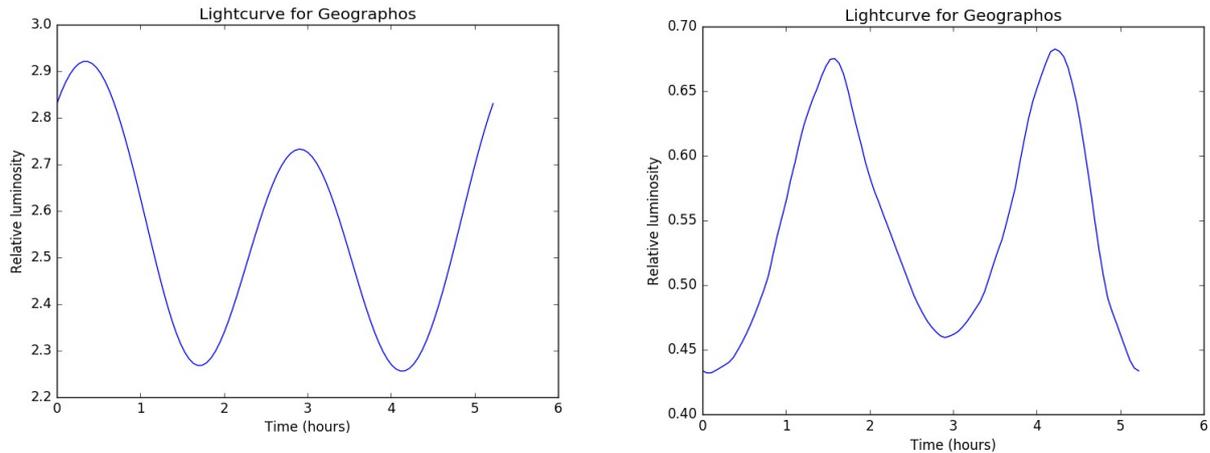

***Fig1***. *Examples of lightcurves for asteroid Geographos, created by our program.*

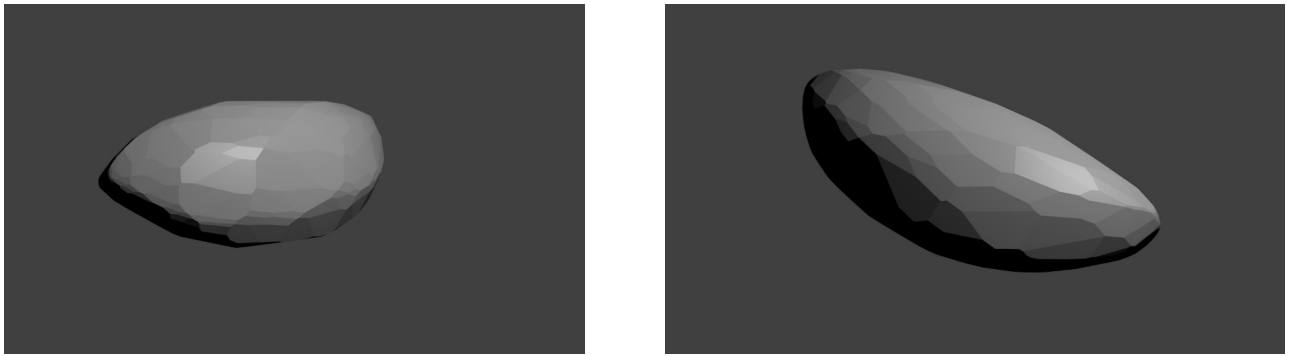

***Fig2***. *Shape of Geographos, taken from DAMIT database.*

## Conclusions

Lightcurve inversion technique is a powerful tool for obtaining asteroid's shapes and rotation parameters. Period of rotation can be obtained with such high accuracy that we are able to detect even lesser deviations caused by YORP effect. At the same time, determination of asteroid's pole and shape has much less precision.

The knowledge about motion and shapes of asteroids gives us important information about genesis and evolution of planets in the Solar system. Our program can be used for analysis of observational data, and, as we hope, serve as the first step to solving inverse problem of asteroid's shape reconstruction.